\documentclass[aps,prl,twocolumn,reprint, superscriptaddress,showpacs,showkeys,10pt]{revtex4-1}
\usepackage{color}
\usepackage{comment}

\usepackage{amsmath}
\usepackage{graphicx}
\usepackage{color}

\newcommand{\dpp}[1]{\textcolor{black}{#1}}
\newcommand{\wxx}[1]{\textcolor{black}{#1}}
\newcommand{\dah}[1]{\textcolor{black}{#1}}

\begin{document}

\title{Overlap junctions for high coherence superconducting qubits}
\author{X. Wu}
\affiliation{National Institute of Standards and Technology, Boulder, Colorado 80305, USA}
\author{J.\ L.\ Long}
\affiliation{National Institute of Standards and Technology, Boulder, Colorado 80305, USA}
\affiliation{Department of Physics, University of Colorado, Boulder, Colorado 80309, USA}
\author{H.\ S.\ Ku}
\affiliation{National Institute of Standards and Technology, Boulder, Colorado 80305, USA}
\author{R. E. Lake}
\affiliation{National Institute of Standards and Technology, Boulder, Colorado 80305, USA}
\author{M. Bal}
\affiliation{National Institute of Standards and Technology, Boulder, Colorado 80305, USA}
\author{D. P. Pappas}\email{David.Pappas@nist.gov}
\affiliation{National Institute of Standards and Technology, Boulder, Colorado 80305, USA}

\date{\today}

\begin{abstract}
Fabrication of sub-micron Josephson junctions is demonstrated using standard processing techniques for high-coherence, superconducting qubits. These junctions are made in two separate lithography steps with normal-angle evaporation. Most significantly, this work demonstrates that it is possible to achieve high coherence with junctions formed on aluminum surfaces cleaned in situ with Ar milling before the junction oxidation. This method eliminates the angle-dependent shadow masks typically used for small junctions. Therefore\dah{,} this is conducive to \dah{the} implementation of typical methods for improving margins and yield using conventional CMOS processing. The current method uses electron-beam lithography and an additive process to define the top and bottom electrodes. Extension of this work to optical lithography and subtractive processes is discussed.

\end{abstract}

\pacs{03.67.Lx}
% 03.67.Lx Quantum computation architectures and implementations

\maketitle

Superconducting devices \dah{implemented as quantum bits (qubits)} are among the leading candidates for building quantum computers. %The key element 
\dah{Key elements} in all types of superconducting qubits %is Josephson junction, which is the non-linear element in the entire superconducting circuitry. 
\dah{are Josephson junctions, which are the non-linear elements} in the superconducting circuitry. This non-linearity separates the two lowest energy levels from higher excitations, %hence forms a two level 
\dah{forming a two-level} system as the physical qubit. Coherence times of superconducting qubits have been increased significantly in both 2D and 3D %geometry (10$\sim$100 $\mu$s)
\dah{geometries ($\sim$10-100$\mu$s)}~\cite{3DQubitsPaikPRL2011,3DRigettiPRB2012,XmonChenPRL2014,SuspendedQubitsChuAPL2016,LossDialSST2016}. These relatively long coherence times, combined with fast, high-fidelity gate schemes, %enable 
\dah{have enabled} the demonstration of quantum error detection \dah{with superconducting devices}~\cite{CRGateChowPRL2011,RIPGateSheldonPRA2016,AdiabticGateMartinisPRA2014}.

While the design and fabrication for \wxx{various other elements that form quantum circuits}, i.e., resonators, shunt capacitors, and inductors, have been well studied and brought into line with standard cleanroom techniques, the preparation of the non-linear Josephson junction is still typically conducted separately %in 
\dah{on a} device-by-device basis. In general, \wxx{low participation ratios from both the} Josephson junction and it's immediate surroundings are essential to the success of present-day superconducting qubits~\cite{SuspendedQubitsChuAPL2016,EpiQubitsWeidesAPL2011}. This goal is typically achieved by shrinking \dah{the} junction size. These \dah{low-loss} junctions have predominantly been fabricated using \dah{a} multi-angle shadow-evaporation (SE) technique, because it naturally yields small structures in a single step process and works well enough for demonstrations of small-scale circuits~\cite{DolanJJ,ManhattanJJ}. SE is also convenient in that the oxidation of \dah{the} base electrode is conducted in-situ on the as-deposited film and, \dah{then} immediately covered by the top electrode. 

\begin{figure*}
\centering
\includegraphics[trim= {0 0 1cm 0}, clip, width = 15cm]{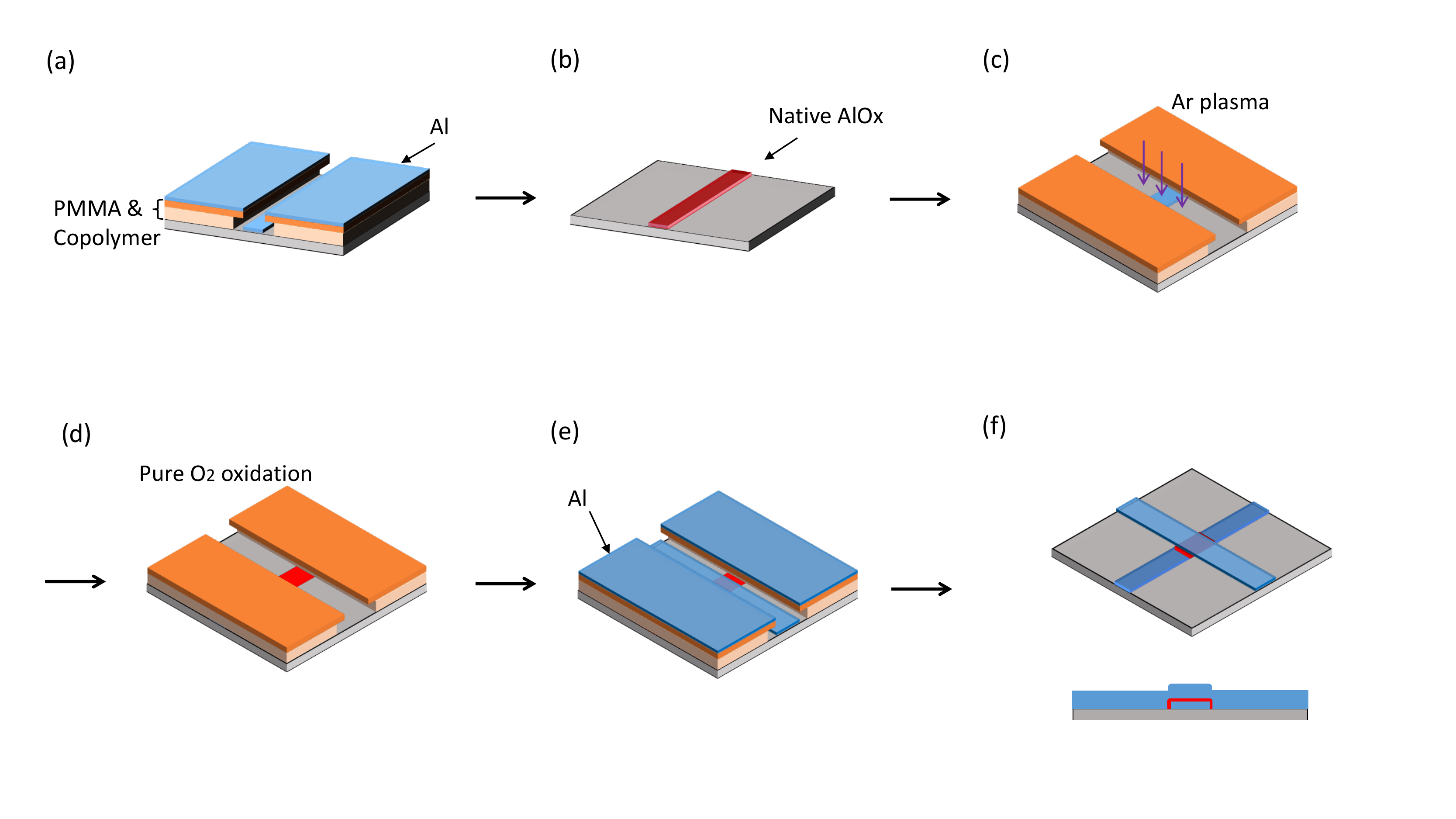}
\caption{Fabrication process flow of overlap junction. 
	(a) 3D view of the device after the first evaporation of Al (bottom electrode). 
	(b) After the metal liftoff, native oxide immediately forms on Al.
	(c) The second lithography defines the pattern of top electrode. Ar RF plasma \dah{cleaning} is performed to remove native oxide on the surface of bottom electrode.
	(d) Low pressure, room temperature oxidation is used to form tunnel barrier.
	(e) Second evaporation of Al to complete the tunnel junction. 
	(f) \dah{Completed overlap junction after the metal-liftoff process}.
}
\label{fig1}
\end{figure*}

%As quantum circuits scale up, 
\dah{To satisfy the requirement of scalability of quantum circuits}, it is becoming critical to bring the junction fabrication step in line with standard fabrication techniques. %to ensure scalability. 
This is difficult with the  angle-dependence of \dah{the} SE technique because it limits the wafer size for preparing junctions with tight margins. One possible avenue is to use overlap junctions, as shown in Ref. \cite{OverlapJJ}, 
where the two electrodes of Josephson junctions are prepared in separate steps. The coherence time from Ref.~\cite{OverlapJJ} is predominately limited by its lossy shunt capacitor and the relatively large junction. Therefore the intrinsic loss of junctions made with the overlap technique could not be evaluated. In this paper, we \dah{describe} the fabrication of %concentric transmon qubit~\cite{Concentric} with small overlap junctions and long coherence times of this qubit is measured. 
\wxx{an overlap-junction}, concentric-transmon qubit~\cite{Concentric} \dah{that exhibits} long coherence times.

Details of the process to form \dah{the} Al/AlO$_{x}$/Al tunnel junction are illustrated in Fig.~\ref{fig1}. Due to the small size of the junction ($\sim$100~nm), we used a standard PMMA/Copolymer (50~nm/100~nm) bilayer resist and electron-beam lithography to pattern the overlap junctions. As shown in Fig.~\ref{fig1}(a), the base electrode is prepared by evaporating Al ($\sim 20$~nm thickness) from an \dah{electron-beam deposition} source. %As we take 
After taking the sample out of the %evaporator
\dah{vacuum chamber, a} native oxide forms immediately on the surface of \dah{the} Al [Fig.~\ref{fig1}(b)]. 
A second lithography is performed to define the top electrode before the sample is loaded into the evaporator again. The tunnel barrier is formed by %an Argon RF plasma clean 
\dah{making use of an argon RF-plasma cleaning step (50~W, 10~mTorr)} followed by room temperature oxidation of the base electrode. 
During this step, we fix the oxidation pressure at 150~mTorr for a short amount of time, typically \dah{$\sim$1-5} minutes. Notably, this oxidation time is an order of magnitude shorter \dah{than} the oxidation time used in the SE technique, of which oxidation happens on fresh deposited Al. This is presumably because the Ar RF plasma not only etches away the native oxide %forms on 
\dah{formed on the} Al surface, but also roughens the Al surface, hence \dah{increasing the rate of oxidation.} Finally, the top Al electrode ($\sim 40$~nm thickness) is \dah{deposited, depicted} in Fig.~\ref{fig1}(e). Figure~\ref{fig1}(f) shows a complete overlap junction after \dah{the metal-liftoff process}. 

We measure \dah{the room-temperature} resistance $R_{n}$ of the junctions made with this method and find \dah{that} the relations \dah{between} $R_{n}$, oxidation time~$t_{\mathrm{ox}}$ %oxidation pressure~$P_{\mathrm{ox}}$ 
and junction area $A$ follow the empirical formula (Eq.1), which is consistent with Ref.~\cite{JJ_Rn}.
\begin{equation}
R_{n} \propto \sqrt{t_{\mathrm{ox}}},~A^{-1} 
\end{equation}

From our measurements, we find this relation holds well for junction \dah{sizes} greater than 0.01~$\mu$m$^{2}$. We believe this lower limit on junction size is due to the process bias of the specific resist used and can be easily mitigated. More importantly, because our method uses normal-angle evaporation, high junction uniformity across a larger area can be achieved. For example, substrate rotation can be implemented during the evaporation to achieve film homogeneity, which is a standard technique used throughout the industry. This overlap method can also be modified for compatibility into a standard processing flow. For example, a subtractive process will yield similar structures given that the top and bottom electrodes are  grown, defined, and patterned using \dah{sputter deposition}, optical lithography, and etching, respectively. Smaller dimensions can also be achieved \dah{via etching}, while some extra attention will be required to avoid redepostion on the edges of the junction, a standard technique in magnetic tunnel junction fabrication~\cite{JunctionEtchingYangJJAP2015}.

\begin{table}
\begin{tabular}{ | p{5cm} | p{3cm} | }
\hline
Qubit frequency ($f_{q}$) & ~$5.647$~GHz \\ \hline
Despersive shift ($\chi$) & ~$1.35$~MHz\\ \hline
Transmon Nonlinearilty ($\alpha$) & ~$262.5$~MHz\\ \hline
Qubit relaxation ($T_{1}$) & ~$34.3~\mu$s\\ \hline
Qubit dephasing ($T_{2}^{*}$) & ~$22.5~\mu$s\\ \hline
Spin echo ($T_{2}^{E}$) & ~$31.4~\mu$s\\
\hline
\end{tabular}
\caption{Qubit Parameters}
\label{table:1}
\end{table}

\begin{figure}[h]
\includegraphics[trim = {7cm 0 7cm 0},clip, width = 9cm]{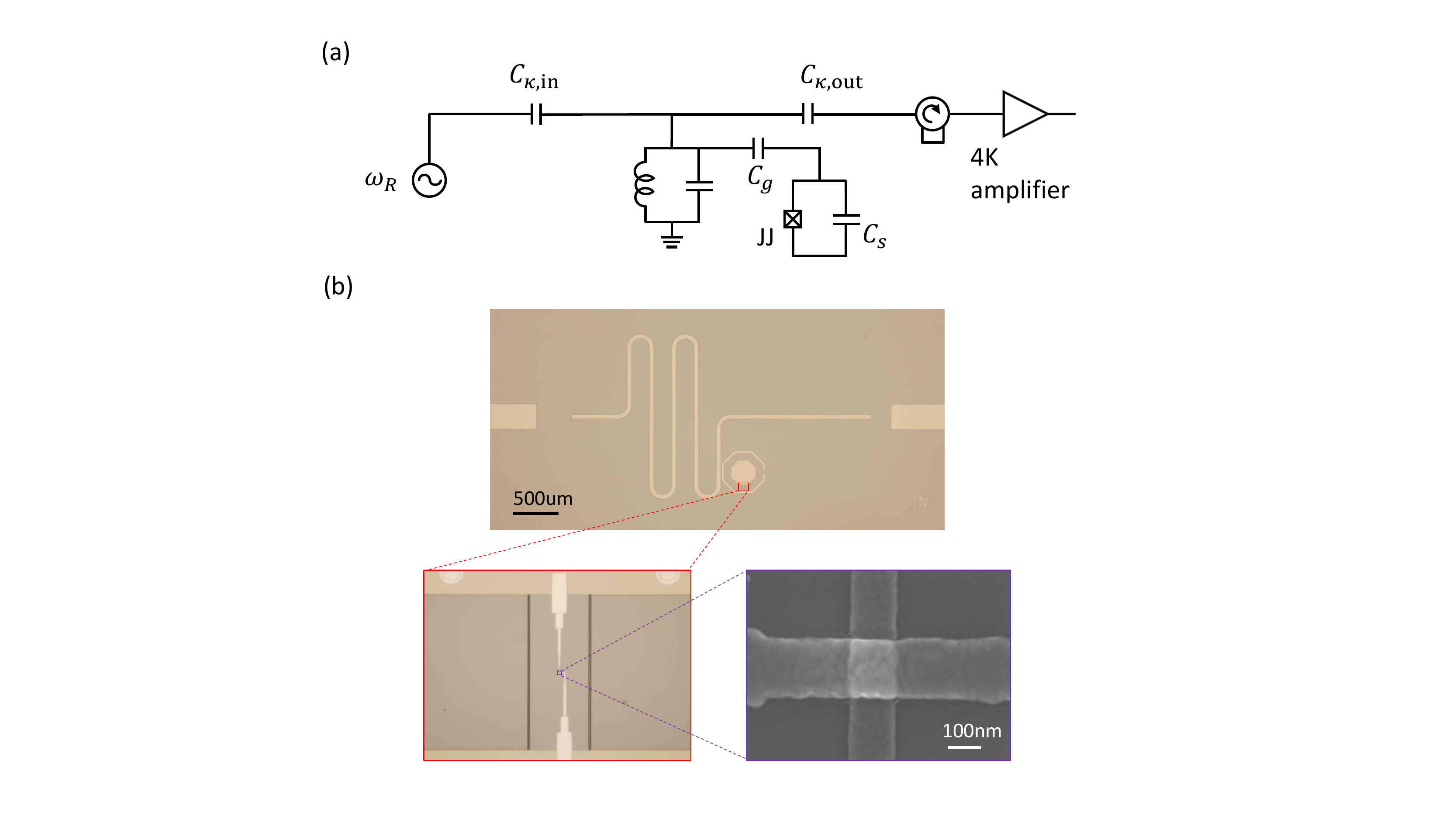} 
\caption{Circuit schematic and micrographs of the qubit device. 
	(a) Circuit schematic showing the concentric tranmon qubit, coupled to a single resonator. The qubit-cavity coupling is governed by $C_{g}$. 
	(b) Optical \dah{and SEM images} of our transmon qubit device. The resonator is realized in microstrip geometry with a measured frequency $\omega_{R}/2\pi = 6.48$~GHz and line width $\kappa/2\pi = 1.1$~MHz.}
\label{fig2}
\end{figure}

\begin{figure}[h]
\includegraphics[trim={8cm 0 4cm 0}, clip, width = 10cm]{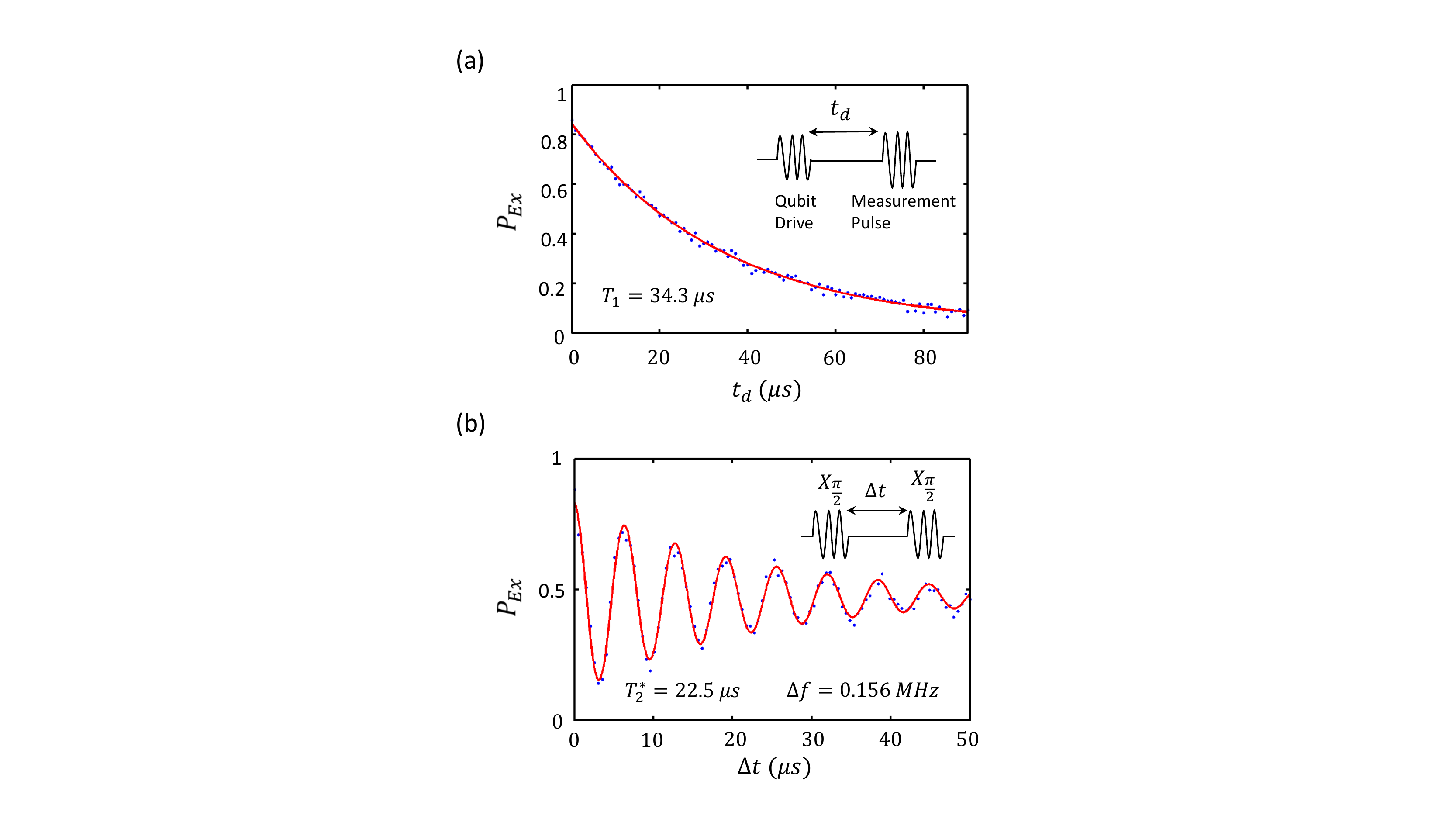}
\caption{ Measured qubit coherence times. 
	(a) Excited State probability as a function of measurement delay $t_{d}$. $T_{1}$ = 34.3$\mu$s extracted from fitting data with exponential decay. 
	(b) Excited state probability as a function of time $\Delta t$ between two $\pi/2$ microwave pulses. $T_{2}^{*}$ = 22.5$\mu$s is extracted from fitting the data with an exponentially damped sinusoid of frequency $0.156$~MHz. }
\label{fig3}
\end{figure}

\dah{To form the large-scale features of our device, such as the resonator and shunt capacitor of the transmon qubit, we use thin-film NbTiN (~35 nm). The NbTiN is grown at 500$^{\circ}$C with reactive sputter deposition and exhibits low loss}~\cite{TiN, TiN_qubit}.
%We use thin film of NbTiN ($\sim$35~nm) to form the big features of our device, such as the resonator and shunt capacitor of transmon qubit. This NbTiN film is grown at 500$^{\circ}$C with reactive sputtering and is extremely low loss. We measured $Q_{i}$ of tens millions at high power and $\sim$800k at low power from lumped resonators made with this material in microstrip geometry. 
\dah{A reactive ion etch} with SF$_{6}$ is used to pattern our device, except \dah{in} the junction area, as shown in Fig.~\ref{fig2}(b). The NbTiN in the junction area is removed \dah{using a} wet etch because it does not attack Si, \dah{leaving} a smooth surface for patterning \dah{the} overlap junctions. The chemical used in this wet etch is NH$_{3}$OH/H$_{2}$O$_{2}$/H$_{2}$O (1:2:5) and the solution needs to be heated above 60C for this etching process.

The concentric transmon design was chosen as a test qubit because of low radiation loss~\cite{Concentric}. The qubit consists of a single overlap junction shunted by a circular capacitor. Unlike the original design, there is a slit in the outer ring to avoid flux trapping. 
Figures~\ref{fig2}(a) and \ref{fig2}(b) show the schematic of our measurement setup and images of our device. The qubit is capacitively coupled to a microstrip $\lambda/2$ resonator ( $\omega_{R}/2\pi = 6.48$~GHz) with \dah{a coupling} strength $g/2\pi=$69~MHz. %Qubit parameters measured from this device is shown in the Table~\ref{table:1}. We measured relaxation ($T_{1} = $34.3~$\mu$s) and decoherence times ($T_{2}=$22.5~$\mu$s) for this qubit. Qubit excited state decay curves are shown in Fig~\ref{fig3}. 
\dah{The qubit parameters measured from this device are shown in Table~\ref{table:1}, and the qubit excited state decay curves are shown in Fig~\ref{fig3}.}
\wxx{The measured relaxation time and decoherence time are $T_{1} = $34.3~$\mu$s and $T_{2}^{*}=$22.5~$\mu$s, respectively. The Purcell limit due to qubit-cavity coupling is 47$\mu$s~\cite{Purcell}, therefore we believe our measured $T_{1}$ could be Purcell limited. We used two methods to measure the decoherence time, which are the ``Ramsey'' time ($T_{2}^{*}$) and the ``Echo'' time ($T_{2}^{E}$)~\cite{T2,Hahn_echo}. The pulse sequence for measuring $T_{2}^{*}$ is shown in the inset of Fig~\ref{fig3}(b). To measure $T_{2}^{E}$, a $\pi$-pulse is inserted half-way between the $\pi/2$ pulses. The $\pi$ pulse reverses the direction of dephasing for the second half of the waiting time ($\Delta$t), thus it ``echoes'' out slow drifts in the qubit frequency. We measured $T_{2}^{E}~\geq~T_{2}^{*}$, suggesting that the qubit is subject to some low-frequency noise that the echo successfully cancels out. However, since $T_{2}^{E}~\leq~T_{1}$, this shows that there is also some decoherence due to high-frequency noise, that the echo cannot remove.}
The fact that qubits made with overlap junctions \dah{have} long coherence times shows that Josephson junctions made with separate steps do not introduce extra loss to the superconducting quantum circuit.

In conclusion, we presented an alternative method of fabricating nano-scale Josephson junctions for superconducting qubits. This method only involves normal-angle evaporation, has no angle dependence, which makes it compatible with large scale fabrication process. We also demonstrated that a 2D transmon qubit made with overlap junctions still has \dah{long} coherence. This opens up the possibility of multi-step fabrication of Josephson junction based qubits.

\begin{acknowledgments}
\dpp{This work was supported by the Intelligence Advanced Research Projects Activity (IARPA) LogiQ Program and the NIST Quantum Based Metrology Initiative. This work is property of the US Government and
	not subject to copyright.}
\end{acknowledgments}
\bibliography{OverlapJunction}
\bibliographystyle{apsrev}

\end{document}